\documentclass[preprint,12pt,3p]{elsarticle}
\usepackage{lipsum}

\makeatletter
\def\ps@pprintTitle{%
	\let\@oddhead\@empty
	\let\@evenhead\@empty
	\def\@oddfoot{}%
	\let\@evenfoot\@oddfoot}
\makeatother

\makeatletter
\renewcommand{\MaketitleBox}{%
	\resetTitleCounters
	\def\baselinestretch{1}%
	\begin{center}
		\def\baselinestretch{1}%
		\Large \@title \par
		\vskip 18pt
		\normalsize\elsauthors \par
		\vskip 10pt
		\footnotesize \itshape \elsaddress \par
	\end{center}
	\vskip 12pt
}
\makeatother

\usepackage{amssymb}
\usepackage{abstract}

\begin{document}
	\begin{frontmatter}
\title{ Superconducting properties of ultrapure niobium}
\author{N. E. Alekseevskiy(1), V. I. Nizhankovskiy\footnote{e-mail: v.nizhan@gmail.com}(1) and K.-H. Bertel(2)}
\address{((1)Institute for Physical Problems, Moscow, Russia,\\ (2)Central Institute of Solid State Physics and Material Science, Dresden, Germany)}
\end{frontmatter}
\begin{abstract}
		The results of determination of critical magnetic fields, magnetization curves and critical currents of niobium samples of different purity are presented. It is shown that ultrapure niobium near $T_c$ is a superconductor of the first type and becomes a superconductor of the second type with decreasing temperature due to the temperature dependence of the Ginzburg-Landau parameters. The dependence of the hysteresis of the magnetization curves of massive samples on the surface state is investigated. A comparison of the superconducting parameters with the parameters of the electron spectrum of niobium is carried out. The dependence of the critical current of niobium wires on the longitudinal magnetic field agrees with the assumption of a force-free current distribution. 
\end{abstract}

\section*{Introduction}
Niobium is known to have a number of interesting physical properties. For example, it is apparently the only pure metal that exhibits the properties of a type-II superconductor. However, there are significant discrepancies between the theory of type-II superconductors and the experimental data obtained on niobium. In particular, the experimentally observed behavior of the magnetization curve in the region between the upper and lower critical fields, as well as the absolute value and temperature dependence of the Ginzburg-Landau parameters, do not correspond to theoretical concepts. Attempts to reduce these discrepancies by increasing the purity of niobium \cite{1, 2} or by extrapolating the properties of niobium alloys with tantalum or molybdenum to an ideally pure metal \cite{3} have proven unsuccessful. Unlike niobium, similar experiments on tantalum \cite{4} show that the purest samples of this metal behave like type-I superconductors, while contaminated ones are similar to type-II superconductors.

Purity of metallic samples is characterized by the residual resistance ratio $\alpha = R(300\,\rm{K})/ \\R(0\,\rm{K})$\footnote{Since niobium is a superconductor, the measurements of the value of $R(0\,\rm{K}$ were carried out in a longitudinal magnetic field $H\geq 8$ kOe.}. Niobium in cited above work \cite{1} had  $\alpha=1,400$, in work \cite{2} $\alpha=2,000$ and in work \cite{3} $\alpha=1,100$. To study the superconducting properties and the Fermi surface of niobium on the purest samples, it is necessary that the value of $\alpha$ be greater than 10,000.

This publication presents results obtained many years ago on ultrapure single crystals and wires of niobium and published in Russian \cite{5}. Because interest to superconductivity of pure niobium has been revived recently \cite{6}, it seems reasonable make old results obtained on uniquely pure samples available in English translation. 

\section{Experiment}
\subsection{Samples}
The complex and rather complicated purification process consisted of five steps was developed in Dresden \cite{7,8}. It comprised the combination of liquid-liquid extraction, chlorination and thermal
decomposition of NbCl$_5$, followed by electron beam float zone melting, decarburization and degassing treatments. A review of the efforts directed to the preparation of ultra-high-purity niobium is given in \cite{9}

Critical temperatures, dependencies magnetization on the field and critical fields were measured on single crystal samples of different orientation and purity. Samples were made from rods with a diameter of 5-6 mm. To obtain the shape of an ellipsoid the rod was melt and stretched from both ends to form a narrowing with the minimum possible diameter in the zone melting. After annealing, the sample was separated at the thinnest point. Length of samples was about 2 cm. Photo of the typical single crystal sample is shown in the insert of Figure 1.

 Table I shows the orientations of the sample axis relative to the crystal axes, resistances ratio and the second critical field extrapolated to $T=0$  for massive single crystal samples on which the magnetization dependencies on the external magnetic field were measured.
\begin{center}
	Table I\\ 
	\medskip  
	\begin{tabular}{c|c|c|c|c|c}
		\hline
		\hline
		& Nb1 & Nb2 & Nb3 & Nb4 & Nb5 \\
		\hline
		orientation & [100] & [100]	& [110] & [111] & [123]\\
		\hline
		$\alpha = R(300\,\rm{K})/R(0\,\rm{K})$ & 36,500 & 61,000 & 34,000 & 52,000 & 28,000\\
		\hline
		$H_{c2}(0)$, kOe & 3.94 & 3.92 & 4.00 & 4.27 & 4.10\\
		\hline
	\end{tabular}
\end{center}

Besides massive single crystals, wire polycrystalline samples were prepared. Their parameters are given in Table II.

\subsection{Devices}
The magnetometer consisted of two counter-connected coils, in one of which the sample being studied was fixed. To record the magnetization curves, the signal from the coils was fed to the integrating photoelectric microwebermeter F-18T and then to the y-coordinate of the two-coordinate recorder. A signal proportional to the current in the solenoid was fed to the x-coordinate. The temperature was setup from 4.2 to 10 K with an accuracy of 10 mK.

It should be noted that in the first magnetization measurements we tried to use the conventional ballistic method, in which the sample is moved between compensated measuring coils. However, the studied single crystals turned out to be so perfect that even a finger click on the cryostat could lead to a jump in magnetization. Therefore, we had to abandon moving the sample.

Critical current measurements were performed by the pulse method on polycrystalline wires made of pure niobium. 
Pulse setup allowed single sawtooth current pulses of up to 500 A with a duration of from 0.3 to 10 msec.  Experiments were done at $T=4.2$ K in a longitudinal homogeneous magnetic field of up to 20 kOe. The use of a small measuring coil placed on a wire sample made it possible to simultaneously measure the potential difference arising on the sample and record the change in its magnetization.

\section{Hysteresis of magnetization of massive samples }
 
The magnetization curves of all the massive samples had distinct hysteresis but there was no residual moment at $H=0$. Maximum hysteresis was observed on the purest samples with a smooth, undamaged surface. The hysteresis detected could hardly be due to the geometry of the samples because their shape was approximately that of an ellipsoid (see insert in Figure 1). 
It also could not be a consequence of the capture of magnetic flux in the sample volume by lattice defects, since in this case a residual moment should have been observed. The absence of flux capture is confirmed by the shape of the partial hysteresis loops (Figure 1) and a decrease in the hysteresis of the magnetization curves when an alternating current with a frequency of 50 Hz was added to the direct supply current of solenoid that created the magnetic field; the amplitude of the alternating field was $\sim 20$ Oe.

\begin{figure}
	\begin{center}
		\includegraphics[width=0.8\textwidth]{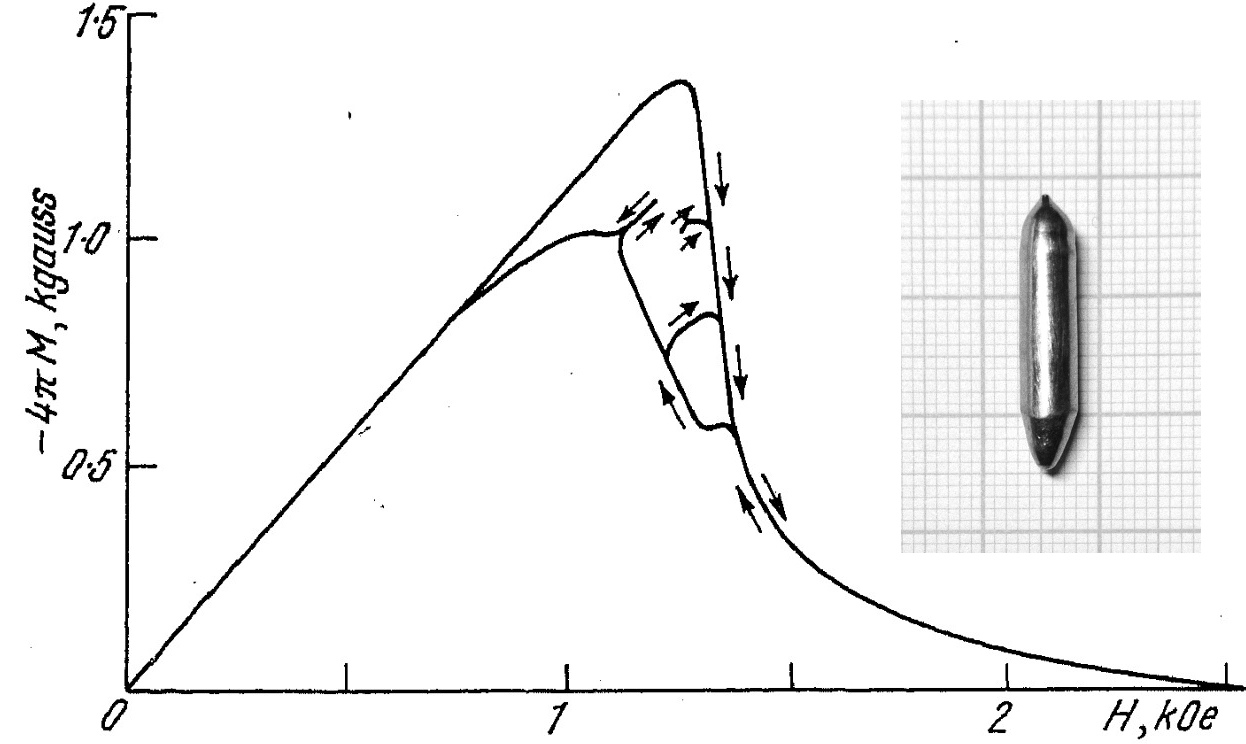}
	\end{center}
	\caption{Magnetization curve of sample Nbl with minor hysteresis loops (copy from trace on a two-coordinate recording chart). Temperature 4.2 K. Surface oxidized by heating in air at 400 $^\circ$C for 4 min. Arrows give direction in which recording pen is moving. Insert: photo of the sample.}
\end{figure}

\begin{figure}
	\begin{center}
		\includegraphics[width=0.8\textwidth]{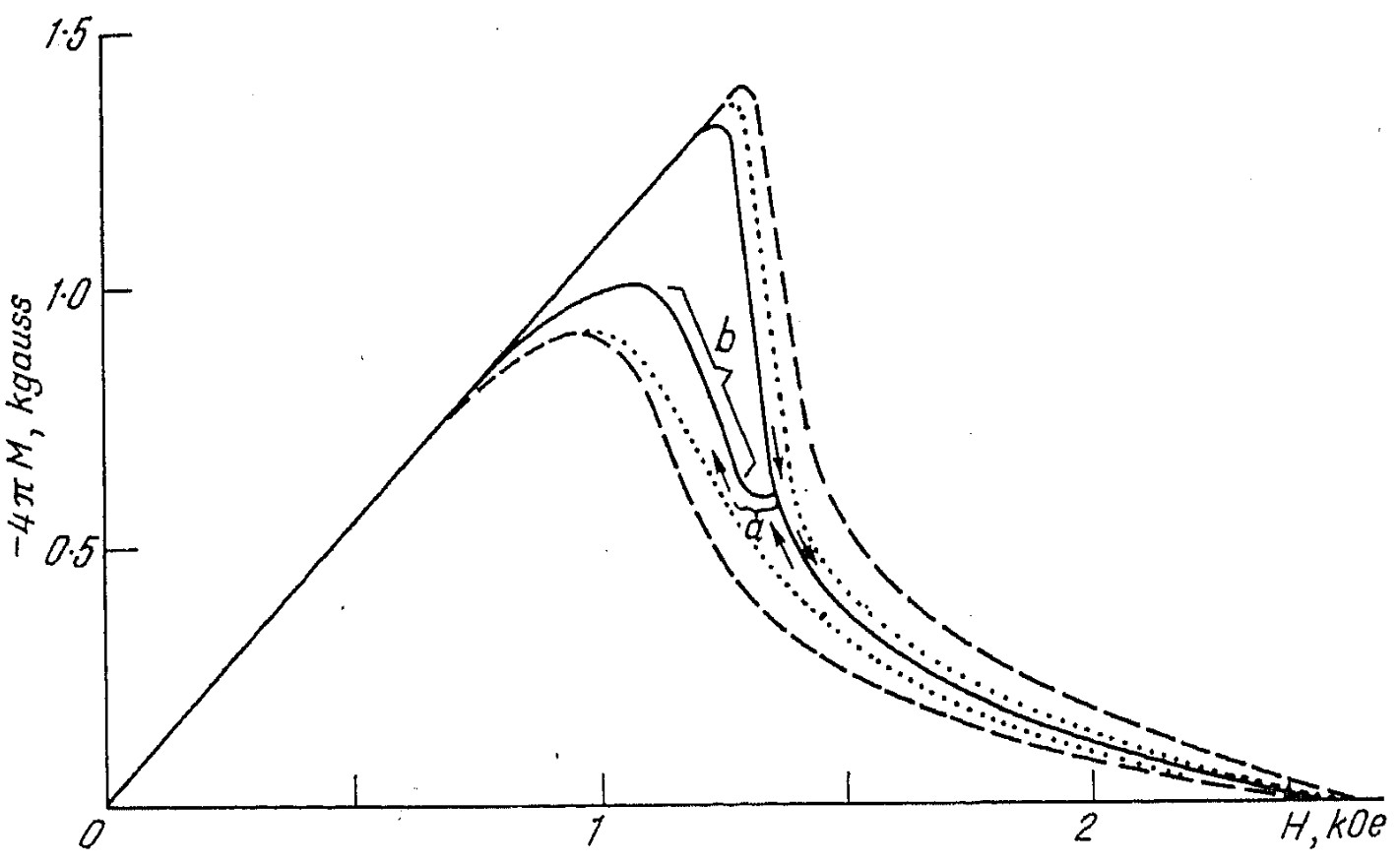}
	\end{center}
	\caption{Influence of surface treatment on magnetization of Nb1  (copy from two-coordinate recordings), temperature 4.2 K: dashed line - before treatment; solid line - after heating in air at 400 $^\circ$C for 4 min; dotted line - after etching. }
\end{figure}
Since irreversibility apparently could not arise in the bulk of the sample, it is natural to assume that its appearance is due to the surface of the sample. To clarify the causes of hysteresis on the Nb1 sample with $\alpha = 36,500$, magnetization measurements were taken after various surface treatments, and the following was established: 1) no significant changes in the magnetization curve occurred with electrolytic deposition of a nickel layer on the surface of the sample; 2) calcination in air at 400$^\circ$C for 1, 2 and 4 min led to a decrease in hysteresis. Magnetization measurements taken after each of the three heat treatments showed that the longer the heat treatment time, the smaller the hysteresis.
After a four-minute calcination, hysteresis was observed only in the vicinity of $H_{c1}$.
If the surface layer was removed by etching, the hysteresis increased again (Figure 2).

The presented facts indicate that the hysteresis above $H_{c1}$ may be caused by surface currents, the occurrence of which is difficult after oxidation of the sample surface \cite{10}. The hysteresis below $H_{c1}$ is divided into two sections. The first (\textbf{\textit{a}} in Figure 2) is caused by either the mutual attraction of the flux filaments  \cite{11,12} or supercooling; the second (\textbf{\textit{b}} in Figure 2) is caused by a surface barrier mechanism similar to that described in  \cite{13}. Section \textbf{\textit{a}}, which was not observed previously, was detected in our measurements due to the high purity of the samples and the finite demagnetizing factor ($n \neq 0$).

\section{Magnetization curves of massive single crystal samples near the critical temperature}
The magnetization curves of sample Nb2 at three temperatures are presented in Figure 3. They are approximately the same for the other pure single crystal samples. Figure 3 shows that when the temperature of the sample is close to $T_c$, $H_{c2}=H_{c1}$ (the finite derivative $dM/dH$ at $H=H_{c2}$ is related with the finite demagnetizing factor),  so ultrapure niobium is a type-I superconductor in the temperature range $(T_c-T)\leq 0.2$ K.

Unfortunately, it is difficult to determine $H_c$, $\kappa_1$ and  $\kappa_2$ with sufficient accuracy from magnetization curves such as those shown in Figure 3 because of their high irreversibility. The hysteresis caused by surface currents could be reduced by oxidizing the sample surface, but over time oxygen could diffuse to a greater depth inside the sample and lower its resistance ratio. In addition, an examination of magnetization curves of samples with specially introduced oxygen showed that an oxygen concentration of $\sim 10$  ppm is sufficient to produce a residual moment. For these reasons, $H_c$, $\kappa_1$ and  $\kappa_2$ were determined only for polycrystalline wire samples, the purity of which was significantly lower than that of massive single crystals.
\begin{figure}
	\centering
	\begin{minipage}{0.47\textwidth}%
		\includegraphics[width=\textwidth]{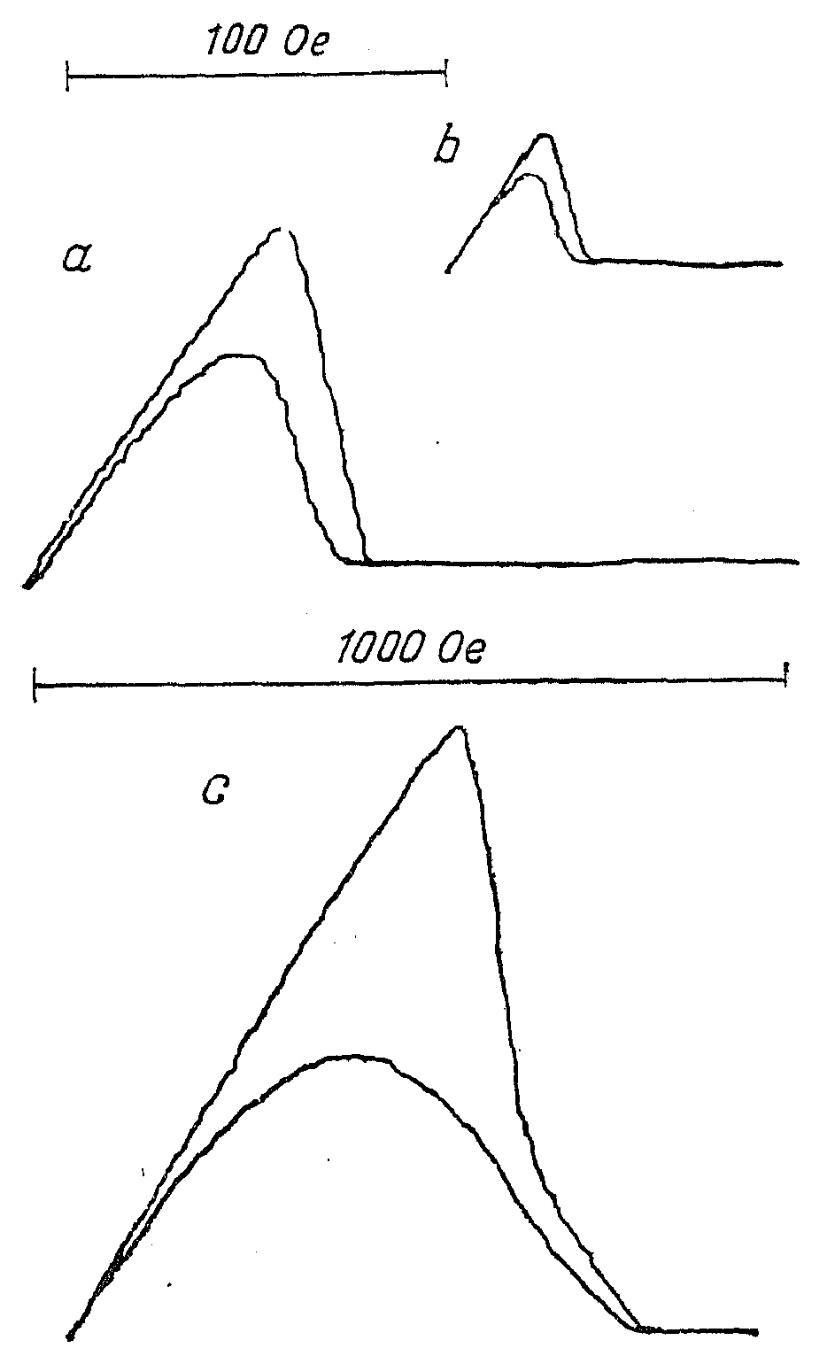}%
		\caption{Magnetization curves of sample Nb2 taken at different temperatures. The scale of 100 Oe relates to curves taken at 9.00 (a) and 9.12 K (b); the scale 1000 Qe is for 7.65 K (c).}
	\end{minipage}%
	\hfill%
	\begin{minipage}{0.47\textwidth}%
		\includegraphics[width=\textwidth]{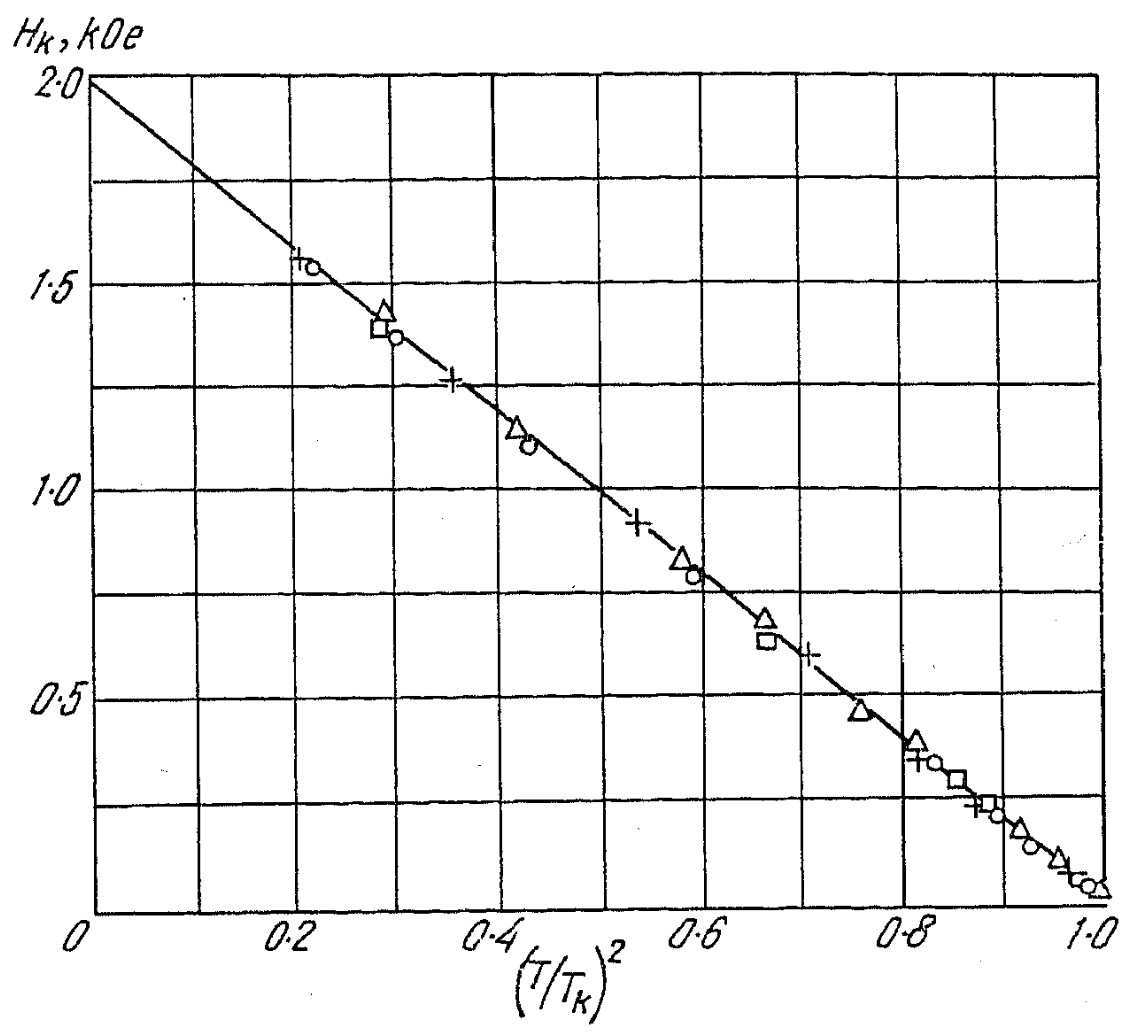}%
		\caption{Temperature dependence of $H_c$: Nb6($\bigcirc$), Nb7($\bigtriangleup$), Nb8($\Box$), Nb9($+$).}
	\end{minipage}%
\end{figure}

\newpage
\section{Dependence of $H_c$, $\kappa_1$, $\kappa_2$ and $\kappa$ on the purity of the wire samples}
\begin{center}
	Table II\\ 
		\medskip  
	\begin{tabular}{c|c|c|c|c|c}
		\hline
		\hline
		& Nb6 & Nb7 & Nb8 & Nb9 & Nb10 \\
		\hline
		$\alpha = R(300\,\rm{K})/R(\,\rm{0\,\rm{K}})$ & 91 & 300 & 5,700 & 9,000 & 15,000\\
		\hline
		$R(300\,\rm{K})\times 100$, Ohm & 3.97 & 3.86 & 4.66 & 3.83 & 3.55 \\		
		\hline
		length, mm & 39 & 42.5 & 38 & 29 & 42 \\
		\hline
		diameter, mm & 0.423 & 0.447 & 0.385 & 0.430 & 0.476 \\		
		\hline
		$H_{c2}(4.2\,\rm{K})$, kOe & 2.91 & 2.735 & 2.68 & 2.68 & 2.67 \\		
		\hline
		$H_{c1}(4.2\,\rm{K})$, kOe & 1.388 & 1.45 & 1.507 & 1.52 & 1.53 \\
		\hline
		$\kappa 1(4.2\,\rm{K})$	& 1.3 & 1.2 & 1.15 & 1.15 & 1.14 \\	
		\hline
		$\kappa 2(4.2\,\rm{K})$	& 1.76 & 1.67 & 1.54 & 1.54 & 1.53 \\	
		\hline
		$H_c(0)$, kOe & 2.0 & 2.0 & 2.0 & 2.0 & 2.0 \\
		\hline
		$\kappa (t=1)$ & 0.85 & 0.748 & 0.705 & 0.703 & - \\
		\hline
	\end{tabular}
\end{center}

The influence of purity on the superconducting parameters
 of the samples was studied on wires whose residual resistance ratio varied between 91 and 15,000 (Table II). In the case of wires the demagnetizing factor $n=0$ and there is no hysteresis related with the attraction of magnetic flux filament. The influence of the surface barrier mechanism is reduced because of the polycrystallinity and small radius of curvature of wire surface.
 
 The thermodynamic critical magnetic field $H_c$ and the Ginzburg-Landau parameters are determined by the following equations \cite{14}: 
 
		\begin{equation}
	\frac{H_c^2}{8\pi}=-\int_{0}^{H_{c2}}MdH;\quad 
	\kappa_1=\frac{H_{c2}}{\sqrt{2}H_c};\quad \kappa_2^2=\frac{1}{2}\biggl[1+\frac{1}{4\pi1.16(\frac{dM}{dH})_{H_{c2}}}\biggr] 
	\end{equation}  
 As we can see from Figure 4, the thermodynamic critical field $H_c$ does not depend on the purity. The temperature dependence of $H_c$ corresponds quite accurately to the quadratic 
\begin{equation}
	H_c(t) =H_c(0)(1-t^2); \quad t=T/T_c
\end{equation}
When $T\rightarrow T_c$ the ratio $H_{c1}/H_{c2}\rightarrow 1$ for the purest samples (Figure 5) and measurements at lower temperatures show that $H_{c2}$ diminishes and $H_{c1}$ grows with increasing purity. 

The temperature dependencies of $\kappa1$ and $\kappa2$ are shown in Figure 6 for samples of different purity. For each sample the experimental points for $\kappa1$ and $\kappa2$ coincide very well at $t=1$, which agrees with the theory \cite{15}. The dependence of parameter $\kappa = \kappa1(1)=\kappa2(1)$ on the residual resistivity may be described by the formula derived in \cite{16}
\begin{equation}
	\kappa=\kappa_0+k\rho\sqrt{\gamma}
\end{equation} 
where $\rho$ is the residual resistivity, Ohm$\cdot$cm; $\gamma$ is the Sommerfeld constant, erg$\cdot$cm$^{-3}$K$^{-2}$. If $\rho \rightarrow 0$, $\kappa\rightarrow \kappa_0 = 0.702 < 1/\sqrt{2}$ (see Figure 6). Thus polycrystalline niobium for which $\alpha>3,000$ is a type-I superconductor at temperatures near $T_c$ and changes to the type-1I when the temperature is lowered because of the temperature dependence of $\kappa1$ and $\kappa2$. 

Fitted to the experimental values of $\kappa$ (bottom line of Table II) the coefficient $k=11.2\cdot 10^3$ was 49\% larger than  calculated in the spherical Fermi surface approximation $k^{sphere}=7.5\cdot 10^3$ \cite{10}. This may point to the essential role of anisotropy of niobium Fermi surface \cite{17,18}.

If we estimate the temperature range in which pure polycrystalline wire samples of niobium will be type-1 superconductors from the temperature dependence of the Ginzburg-Landau parameters, it will be significantly narrower than the temperature range in which massive single-crystal samples behave as type-1 superconductors. It is possible that the large surface-to-volume ratio and imperfection of the surface of   polycrystalline wire samples may cause an increase in $\kappa$.
\begin{figure}
	\centering
	\begin{minipage}{0.47\textwidth}%
		\includegraphics[width=\textwidth]{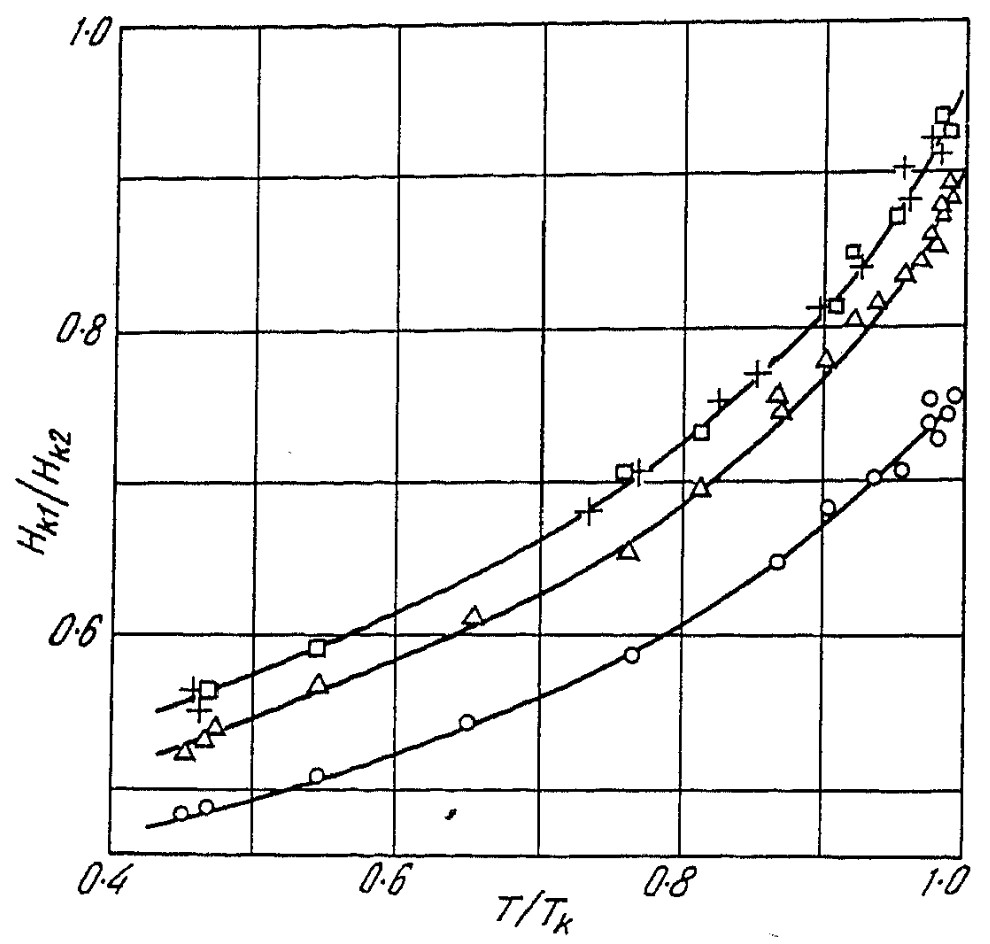}%
		\caption{Temperature dependence of $H_{c1}/H_{c2}$: Nb6($\bigcirc$), Nb7($\bigtriangleup$), Nb8($\Box$), Nb9($+$).}
	\end{minipage}%
	\hfill%
	\begin{minipage}{0.47\textwidth}%
		\includegraphics[width=\textwidth]{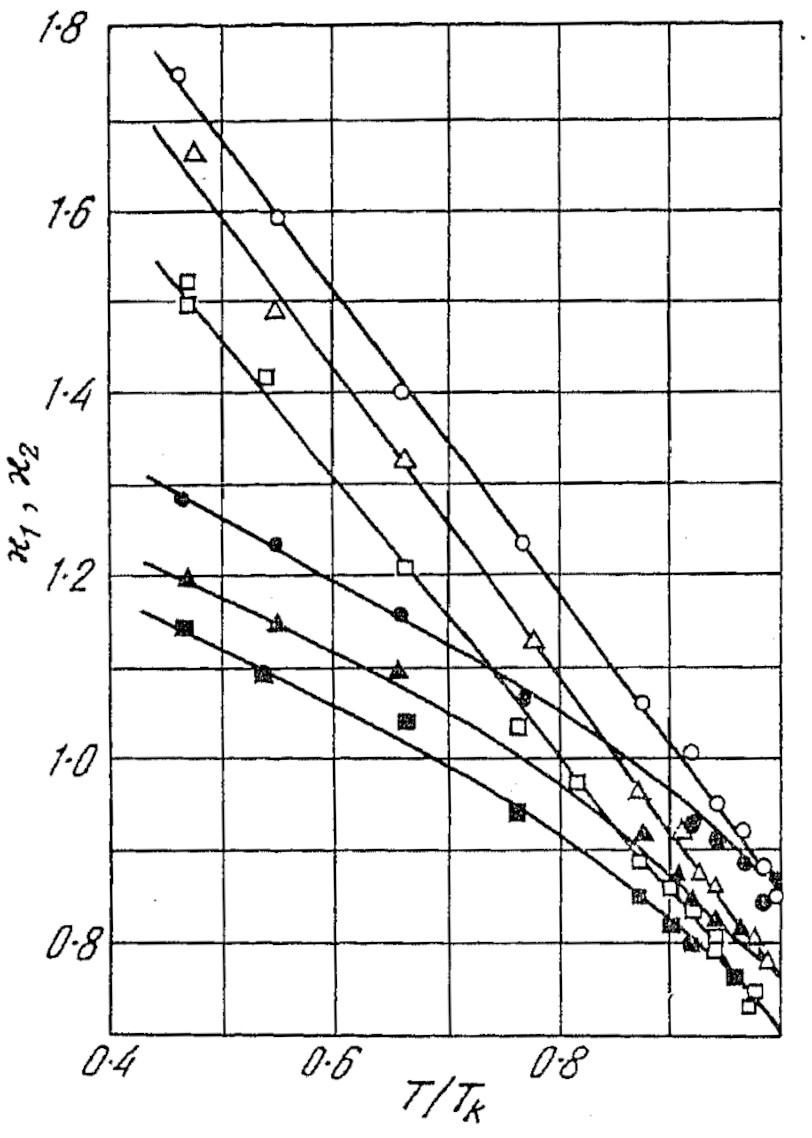}%
		\caption{Temperature dependence of $\kappa1$ (filled symbols) and $\kappa2$ (empty symbols): Nb6($\bigcirc$), Nb7($\bigtriangleup$), Nb8($\Box$).}
	\end{minipage}%
\end{figure}

\section{Anisotropy of the second critical field}

The influence of crystallographic orientation on $H_{c2}$ was studied on pure ($\alpha > 30,000$) samples at temperatures significantly lower than $T_c$. From Figure 7 it is evident that while at $t^2>0.6$ the difference in  $H_{c2}$ values for different orientations lies within the measurement error, at $t^2<0.6$ it can be determined unambiguously. The biggest difference arises between $H_{c2}$ for [100] and $H_{c2}$ for [111]. Table 1 gives the values of $H_{c2}(0)$ of different samples. The following formula was used for the extrapolation
\begin{equation}
	H_{c2}(t)=H_{c2}(0)[1+\eta t^2\ln{t}]
\end{equation}
where the factor $\ln{t}$ describes non-local effects in pure superconductors \cite{19,20}.For metals with a cubic crystal lattice  anisotropy $H_{c2}$ in the local approximation is absent \cite{21}. The coefficient $\eta^{sphere}=0.65$ for a spherical Fermi surface   is very different from our measurement $<\!\!\eta\!\!>\, = 1.6$, so the Fermi surface anisotropy must have a strong influence on the superconducting properties.
\begin{figure}
	\centering
	\begin{minipage}{0.47\textwidth}%
		\includegraphics[width=\textwidth]{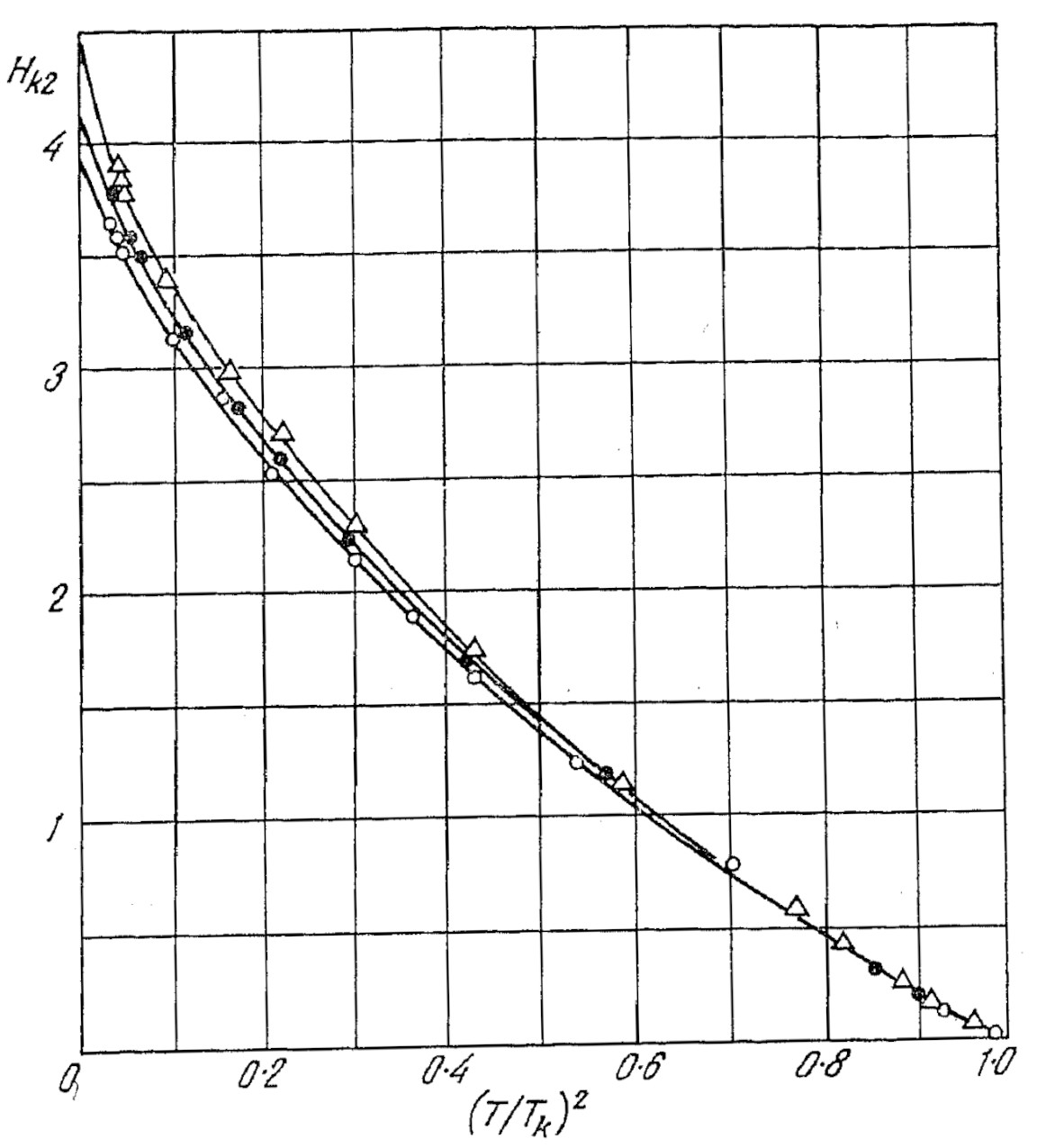}%
		\caption{Temperature dependencies of $H_{c2}$ for single-crystalline samples with different orientations: Nb2[100] ($\bigcirc$), Nb3[110]($\bullet$), Nb4[111]($\bigtriangleup$).}
	\end{minipage}%
	\hfill%
	\begin{minipage}{0.47\textwidth}%
		\includegraphics[width=\textwidth]{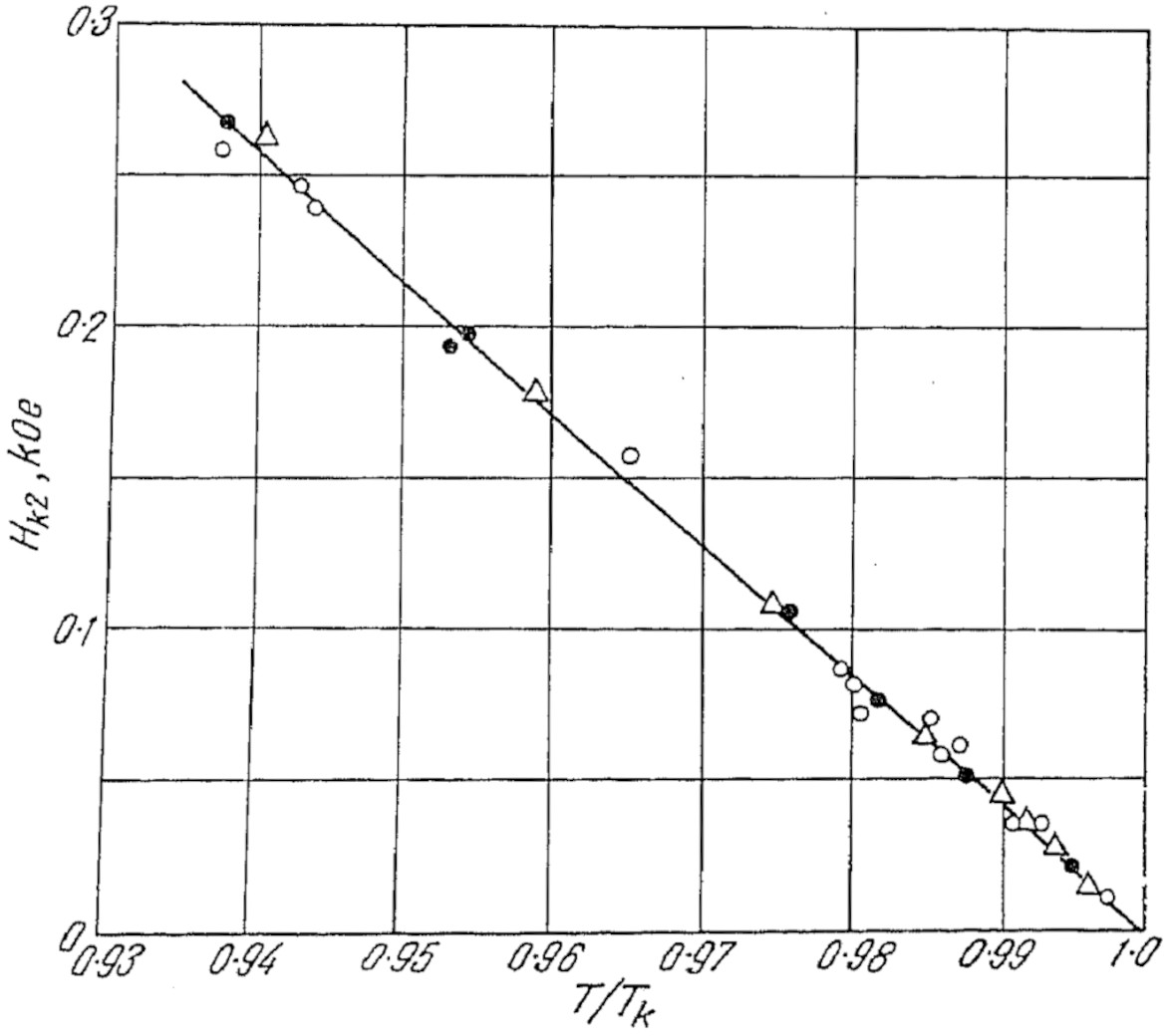}%
		\caption{Temperature dependence of $H_{c2}$ near $T_c$ for single-crystalline samples with different orientations: Nb2[100] ($\bigcirc$), Nb3[110]($\bullet$), Nb4[111]($\bigtriangleup$).}
	\end{minipage}%
\end{figure}

In principle, the anisotropy of $H_{c2}$ can be caused by the anisotropic distribution of impurities, but their concentration in the samples studied is so small that the mean free path $l$ of electrons is several orders of magnitude greater than the coherence length $\xi_0$ (Table III ) and therefore the influence of this mechanism on $H_{c2}$ of the samples we used is unlikely.

The obtained values of the difference $H_{c2}$ are consistent with the data of Williamson \cite{22}, but the absolute values of $H_{c2}$ for our purest samples are 2\% lower.

Average value of $<\!\!H_{c2}(0)\!\!>\,=4160$ Oe.
The slope of the $H_{c2}(t)$ near $t=1$  for all samples has approximately the same value (Figure 8)
\begin{equation}
	-\Bigl(\frac{dH_{c2}}{dt}\Bigr)_{t=1}=4310 \,\rm{Oe}
\end{equation}
Introducing the normalized value
\begin{equation}
	<\!\!h(t)\!\!>\,=-H_{c2}(t)/\Bigl(\frac{dH_{c2}}{dt}\Bigr)_{t=1}
\end{equation} 
we shall get $<\!\!h(0)\!\!> \,=0.97$. This is in good agreement with the $<\!\!h(0)\!\!>\, =0.99$ calculated by Mattheiss from data on the Fermi surface of niobium \cite{17}. The computation \cite{23} for the case of a weak coupling gives the following relation between $<\!\!v_F^2\!\!>$ and $\Bigl(\frac{dH_{c2}}{dt}\Bigr)_{t=1}$:
\begin{equation}
<\!\!v_F^2\!\!>\,=-\frac{6c(2\pi k_BT_c)^2}{7\zeta(3)e\hbar\Bigl(\frac{dH_{c2}}{dt}\Bigr)_{t=1}}= - \frac{2.705\cdot 10^{18}}{\Bigl(\frac{dH_{c2}}{dt}\Bigr)_{t=1}}\,\rm{[cm/sec]}^2 
\end{equation} 
In the case of a strong coupling additional coefficient $1.4(1+\lambda)^2$ appear in Eq.7 \cite{22,24}. Then for $<\!\!v_F^2\!\!>$ we shall have 
\begin{equation}
	<\!\!v_F^2\!\!>\,= - \frac{3.78\cdot 10^{18}}{\Bigl(\frac{dH_{c2}}{dt}\Bigr)_{t=1}}(1+\lambda)^2 
\end{equation} 
According to measurements of the temperature dependence of the de Haas-van Alphen effect \cite{25} $1+\lambda=2.14$ for niobium. Thus $<\!\!v_F^2\!\!>^{1/2}\,=6.3\cdot 10^7$ cm/sec, which agrees very well with Mattheiss’ $6.2\cdot 10^7$ cm/sec \cite{17}.

\begin{center}
	Table III Superconducting parameters of ultrapure niobium. \\ 
	\medskip  
	\begin{tabular}{c|c|c|c}
		\hline
		\hline
parameter & this work  & \cite{26} & \cite{2} \\
		\hline
		\hline
$\Theta_D$, K & - & 275 & - \\ \hline 
$\gamma$, 10$^3$erg/cm$^3$K$^2$ & 7.4 &7.2&7.3\\ \hline
$T_c$, K&9.22&9.19&9.30\\ \hline
		$H_c(0)$, Oe & 1990&1994&1980 \\ \hline
$<\!\!H_{c2}(0)\!\!>$, Oe&4160&-&- \\ \hline
$-\Bigl(\frac{dH_{c2}}{dt}\Bigr)_{t=1}$, Oe& 4310&-&- \\ \hline
$H_{c1}(0)$, Oe& 1860&-&- \\ \hline
$\kappa_0(t=1)$&0.702&-&0.815 \\ \hline
$\kappa_1(t=0)$ &1.48 &-& 1.39 \\ \hline 
$\lambda_L(0)$, nm & 34.3 & - & 33.3\\
\hline
$\lambda_0(0)$, nm&41.4&-&39.7\\ 
\hline
$\xi_0(0)$, nm&47.0&-&39.0\\		
\hline
$2\Delta(0)/k_BT_c$&3.63&3.69&3.62\\
\hline
$<\!\!v_F^2\!\!>^{1/2}$, 10$^7$cm/sec & 6.3&-&-\\		
\hline
$S_F$, 10$^{17}$cm$^{-2}$& 4.9&-&5.15\\		
\hline
n, 10$^{23}$ cm$^{-3}$&2.3&-&1.97\\
\hline
$<\!\!l\rho_0\!\!>$, 10$^{-12}\,\Omega\cdot$cm$^2$&3.1&-&3.75\\
\hline
$l/\xi_0$ for $\alpha>3\cdot 10^4$& $>1,300$&-&-\\
\hline
	\end{tabular}
\end{center}

\section{Critical current}

The destruction of superconductivity by current in type-I superconductors was theoretically considered in the works of London and Landau  \cite{27}.
According to these works, at $I_c =5rH_c$  ($r$ is the radius of the sample, cm; $H_c$ - critical magnetic field, Oe; $I_c$ - critical current, A) the sample goes into an intermediate state, and its resistance increases abruptly to $\frac{1}{2}R_{norm}$  and with a further increase of current asymptotically tends to $R_{norm}$ - resistance of the sample in the normal state.
The experimentally determined behavior of resistance during the destruction of superconductivity by current
corresponds to the theoretical one. If a sample of a type-I superconductor is placed in an external longitudinal magnetic field $H$, then resistance arises when the total field on the surface of the sample is equal to the critical field (Silsbee's hypothesis \cite{28}) and the dependence of the critical current of the sample  on the external field $H$ is a circle.

\begin{figure}
	\centering
	\begin{minipage}{0.47\textwidth}%
		\includegraphics[width=\textwidth]{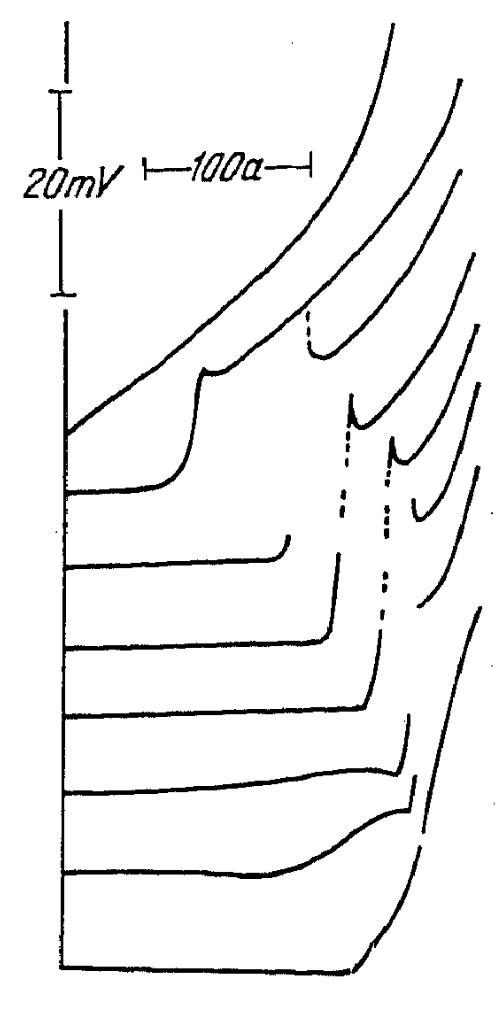}%
		\caption{Copies of V-I oscillograms of the current destruction of superconductivity of sample Nb7 in longitudinal magnetic field. Temperature 4.2 K. Pulse length 1.2 msec. Magnetic field (upwards): 0; 0.96; 1.44; 1.92; 2.16; 2.40; 2.63; 2.87 kOe.}
	\end{minipage}%
	\hfill%
	\begin{minipage}{0.47\textwidth}%
		\includegraphics[width=\textwidth]{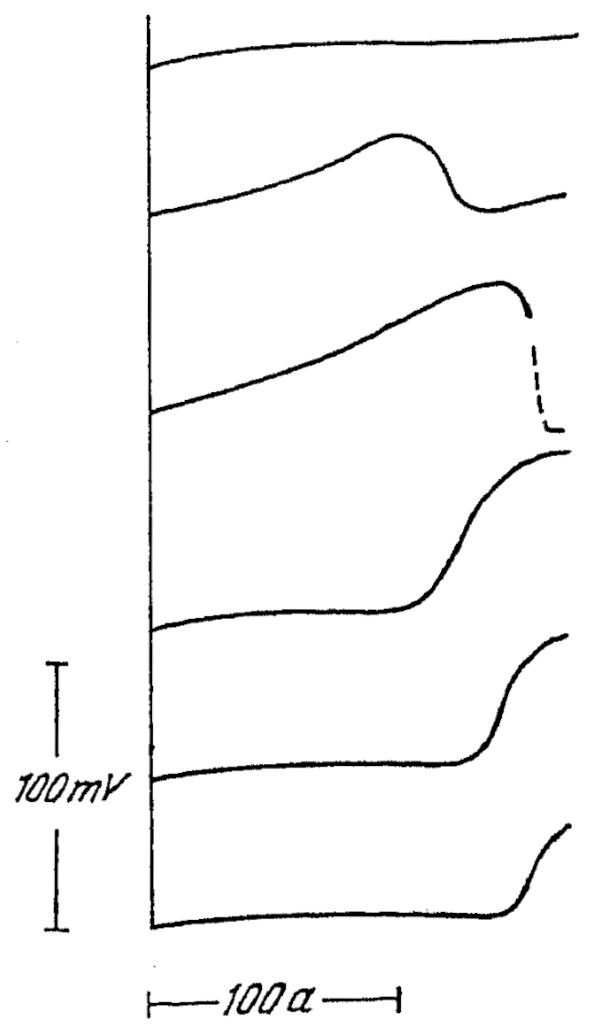}%
		\caption{Copies of oscillograms of the signal taken from
			an ballistic coil wound on a sample, obtained during the destruction	of the superconductivity of the sample
			Nb10 by current in longitudinal magnetic field at $T=4.2$ K;
			duration of current pulses 4.2 msec. Magnetic field
			(from bottom to top): 0.72; 0.96;
			1.20; 216; 2.40; 2.70 kOe. Number of coil turns 5260.}
	\end{minipage}%
\end{figure}

The critical current of type II superconductors depends not only on the external magnetic field. It is largely determined by various mechanisms leading to magnetic flux capture \cite{29}, therefore the theoretical explanation of the experimental results obtained when measuring the critical currents of “dirty“ type-II  superconductors is not always unambiguous. It was of interest to measure the critical currents of high-purity niobium wires, in which magnetic flux capture is almost completely absent.

Figure 9 shows the oscillograms recorded during the measurement of the critical currents of the Nb7 sample in a longitudinal magnetic field. The oscillograms obtained on other samples have approximately the same appearance. At $H = 0$ the voltage signal arises when $I_{c1} =5rH_{c1}$, which corresponds to the transition of the sample to a mixed state. The complete transition to the normal state occurs at a current of $I_{c2}\approx 2.5rH_{c2}$ which is half the critical current determined by the Silsbee rule. Such a relatively low value of the critical current cannot be caused by overheating of the sample. If the sample is in a superconducting state, then the heat release at the current contacts (the resistance of which turned out to be approximately $2\cdot 10^{-6}$ Ohm) gives a temperature increase of the sample of no more than 0.3 K (the estimate is made for the adiabatic regime). 
The value of
$I_{c2}$ can apparently be explained by the force-free distribution of current \cite{30}, in which the current lines are not parallel to the axis of the sample, but are curved in such a way that the local Lorentz force is equal to zero
\begin{equation}
	\textbf{i}\times \textbf{b}=0
\end{equation}

The current line described by the solution of Eq.(9) is a
spiral \cite{30,31}. 

Figures 11 and 12 show the dependencies of the reduced critical current on the external magnetic field for samples with greatly differing purity. With increasing duration of the current pulses (see Figure 11), the agreement with the theory improves. The point, apparently, is that in \cite{23} only stationary states were considered, while in our measurements the current continuously changes. The increase in current can also explain the signal observed at $H<H_{c1}$, corresponding to the flux flow resistance (see Fig. 9). In this case, for each fixed value of the current through the sample, there is a stationary distribution of the flux lines; when the current changes, the distribution changes, which causes a potential difference on the sample.

If $H>H_{c1}$, then the flux flow resistance is small and only the transition to the normal state is recorded. It should be noted that in this case, at $I<I_{c2}$, a paramagnetic signal
 was recorded on the ballistic coil used to measure the magnetic moment(Figure 10). Its appearance is consistent with the force-free distribution of the current \cite{32}.
 
On the oscillograms in Figure 9, one can see a voltage peak that occurs at the transitions to the normal state. Its appearance
is apparently associated with the transition from a spiral force-free current distribution to a conventional uniform one.

\begin{figure}
	\centering
	\begin{minipage}{0.47\textwidth}%
		\includegraphics[width=\textwidth]{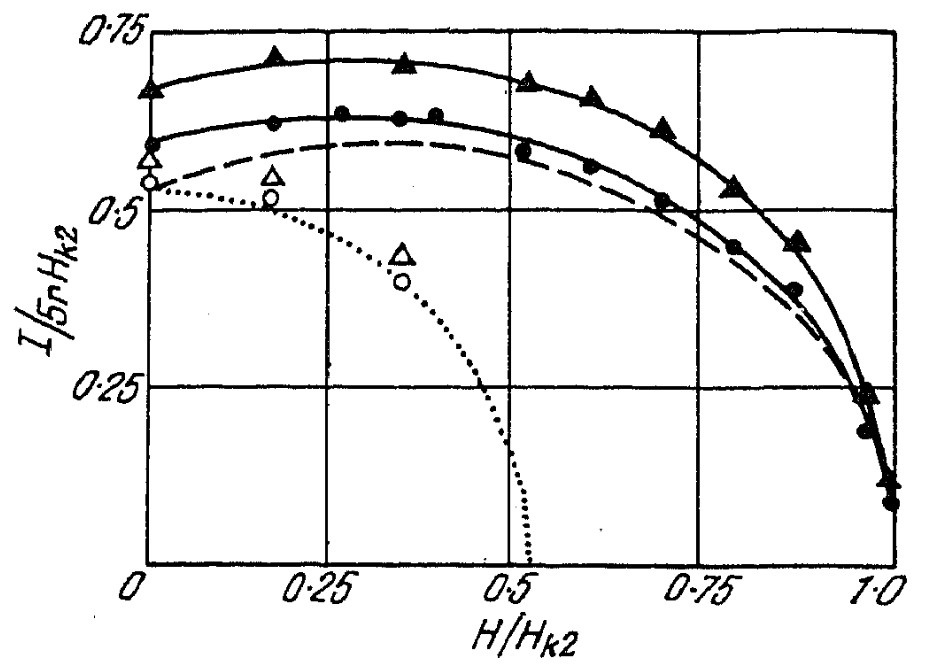}%
		\caption{Dependences of the reduced critical current on the external magnetic field of the Nb7 ($\alpha = 300$) sample for current pulses of 1.2 ($\bigtriangleup, \blacktriangle$) and 4.2 msec ($\circ, \bullet$ ), T—4.29 K. $\circ, \bigtriangleup$ - occurrence of a signal corresponding to the resistance of the flow of current; $\bullet, \blacktriangle$ - transition to the normal state. The dotted line limits the region of Meissner states, the dashed line is the theoretical dependence for a force-free current distribution \cite{23}}
	\end{minipage}%
	\hfill%
	\begin{minipage}{0.47\textwidth}%
		\includegraphics[width=\textwidth]{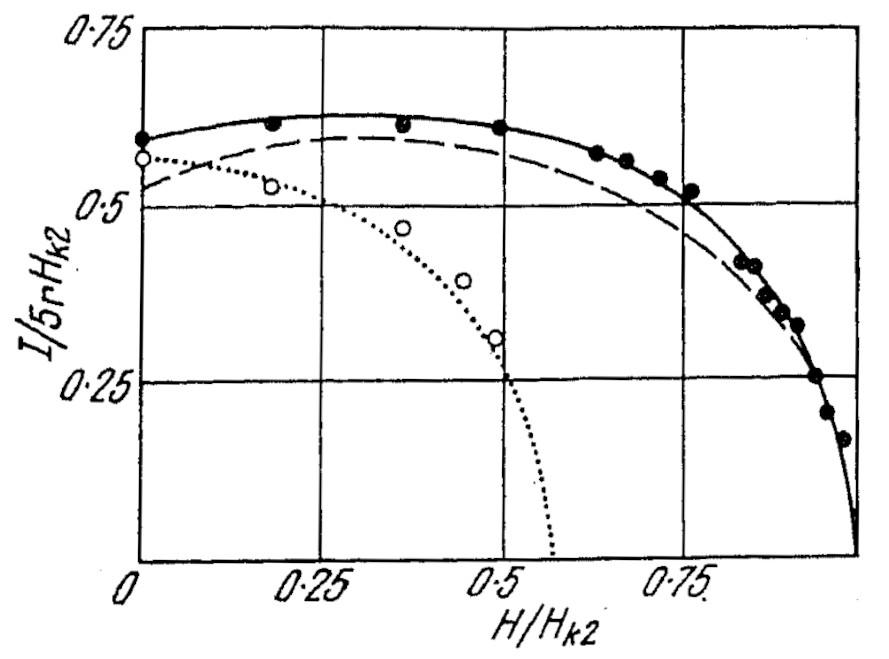}%
		\caption{Dependence of the reduced critical current on the external magnetic field of the Nb10 ($\alpha = 15,000$) sample; $T=4.29$ K; current pulse duration 4.2 msec;
			$\circ$ - occurrence of a signal corresponding to
			the flux flow resistance; $\bullet$ -transition
			to the normal state. The dotted line
			limits the region of Meissner states, the dashed line is the theoretical dependence
			for a force-free current distribution \cite{23}.}
	\end{minipage}%
\end{figure}

According to \cite{31}, a difference in the destruction of superconductivity by current should be observed for superconductors with critical field ratios of $H_{c1}/H_{c2}<0.52$ and $H_{c1}/H_{c2}>0.52$. If $H_{c1}/H_{c2}<0.52$ and
$H<H_{c1}$, then the increase in current first leads to the transition of the sample to a mixed state, and then to a normal state. For the Nb7 sample, the ratio $H_{c1}/H_{c2}\simeq 0.52$ and, indeed, in low fields a transition to a mixed state is observed (at $I>I_{c1}$ a flux flow resistance signal appears).

If $H_{c1}/H_{c2}>0.52$, then the transition to the normal state occurs bypassing the mixed state \cite{31,33}. For the Nb10 sample $H_{c1}/H_{c2}=0.57$ and at $H=0$ the signal corresponding to the flux flow resistance was not detected.
\section*{Conclusions}
1. Ultrapure niobium near $T_c$ is a type-I superconductor.
The temperature range in which extremely pure single crystal niobium remains a type-I superconductor is relatively small
($T_c-T \le 0.2$ K).

2. The hysteresis of the magnetization curves observed on massive single crystal samples depends on the surface state.
The hysteresis above $H_{c1}$ is apparently caused by surface currents, and surface oxidation, preventing their occurrence, suppresses the hysteresis.

3. The values of $<\!\!h(0)\!\!>$ and $<\!\!v_F^2\!\!>^{1/2}$ calculated from our measurements are in good agreement with the results obtained by Mattheis \cite{17}.

4. The results of measuring the critical currents of niobium wires in a longitudinal magnetic field can be explained by the force-free current distribution, in which the current lines are curved in such a way that the current and field are mutually parallel at any point of the sample. An earlier attempt to observe force-free states in type-II superconductors did not yield positive results \cite{34}, which may be due to the large influence of mechanisms leading to magnetic flux capture on the establishment of force-free states. In our samples, according to magnetization measurements, magnetic flux capture is completely absent.

\end{document}